\documentstyle[twoside, epsf]{article} 
 
\input ibvs2.sty 
 
\begin{document} 
\IBVShead{XXXX}{24 Oct 2005} 
 
\IBVStitle{Discovery and disentangling of multi-mode $\delta$-Sct}
\vspace{-12pt}
\IBVStitle{pulsations in the eclipsing binary V994 Her} 
  
\IBVSauth{Sergio Dallaporta$^1$ and Ulisse Munari$^2$}

\IBVSinst{Via Filzi 9, I-38034 Cembra (TN), Italy} 
\IBVSinst{INF Osservatorio Astronomico di Padova, Sede di Asiago, I-36032 Asiago (VI), Italy} 
 
\SIMBADobjAlias{V994 Her}{HD 170314}{HIP 90438} 
\IBVStyp{EA} 
\IBVSkey{photometry} 
\IBVSabs{XXX}
\IBVSabs{XXX}
\IBVSabs{XXX}
\IBVSabs{XXX}
\IBVSabs{XXX}
\IBVSabs{XXX}
\IBVSabs{XXX}

\begintext 

V994~Her (= HD 170314 = BD +24.3425, spectral type A0) has been discovered
as a variable star by the Hipparcos satellite (HIP 90483, $V_T$=7\fmm00,
$B_T$=7\fmm01, $H_P$=6\fmm95) which recognized it as an eclipsing binary
star with the following ephemeris giving times of primary minima:
\begin{equation}
   {\rm Min.\ I = HJD\ }2448501.1239\ \ +\ \  2\fday0830900 \times E
\end{equation}
The brightness is $H_P$=6\fmm93 outside eclipses and $H_P$=7\fmm24 at primary
minimum. V994~Her is located at $\alpha$=18 27 45.89 and
$\delta$=+24 41 50.7 (J2000.0), corresponding to galactic coordinates $l$=52.94 and
$b$=+15.90. The parallax measured by Hipparcos is $\pi=4.14\pm1.16$, for a
distance of $\sim$240~pc.

V994~Her is since long known as a close double. Among others, Abt (1985)
gives for 1952 a P.A.=31.4~deg and a separation 1.93 arcsec, while Ling \&
Prieto (2000) provide for 1999 a P.A.=358.4 deg and a separation 1.20
arcsec. Hipparcos measurements for 1991 gives P.A.=0.7~deg and a separation
1.20 arcsec. From Tycho data, Fabricius \& Makarov (2000) derived
$V_T$=7\fmm16, $B_T$=7\fmm17 for the brighter component (=V994~Her itself, the
eclipsing variable) of the astrometric pair, and $V_T$=9\fmm00, $B_T$=9\fmm24
for the fainter companion 1.2 arcsec away, which could be itself variable in
brightness. Apart from these astrometric measurements and timing of some
eclipses (Borkovits et al. 2002, 2004; Ak \& Filiz 2003), not much else is
known for V994~Her.

We observed V994~Tau in $V$ band (standard Johnson filter) from a private
observatory near Cembra (Trento), Italy, in a similar way to our previous
investigations of V432~Aur (Siviero et al. 2004) or HD~23642 (Munari et al.
2004). The instrument was a 28 cm Schmidt-Cassegrain telescope equipped with
an Optec SSP5 photometer. The diaphragm had a size of 74 arcsec, and the
exposure time was usually 10 seconds. HD\,168957 (HIP\,89975,
$V_J$=6\fmm978, $(B-V)_J$=$-$0.095, spectrum B3V) was chosen as comparison
star and HD\,336061 (TYC 2097-687-1, $V_J$=8\fmm58, $(B-V)_J$=+1$\fmm$37,
spectrum K5; as for HD\,168957, Johnson's $B_J$, $V_J$ are derived from
Tycho's $B_T$, $V_T$ values following Bessell 2000 transformations) as a
check star. The comparison has been measured by Hipparcos 187 times and
found constant. We have measured it against the check star 23 times in
different nights and found $V_J$=8\fmm 58 with $\sigma_V$=0.01 mag, thus pretty
well confirming the absence of variability.

\IBVSfig{8cm}{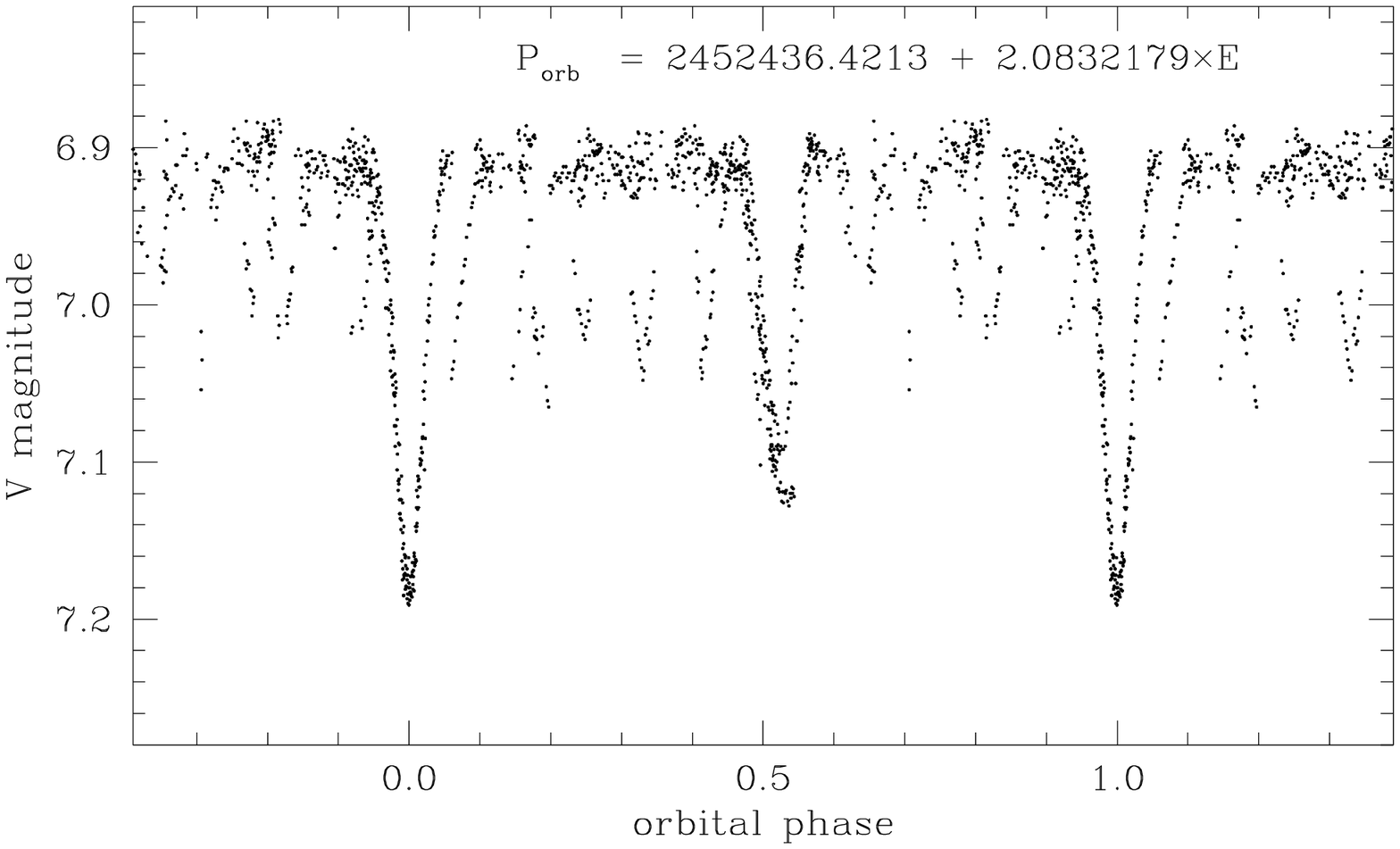}{Our V-band photometry of V994 Her folded onto the orbital ephemeris given in Eq. (2)}

\IBVSfig{12cm}{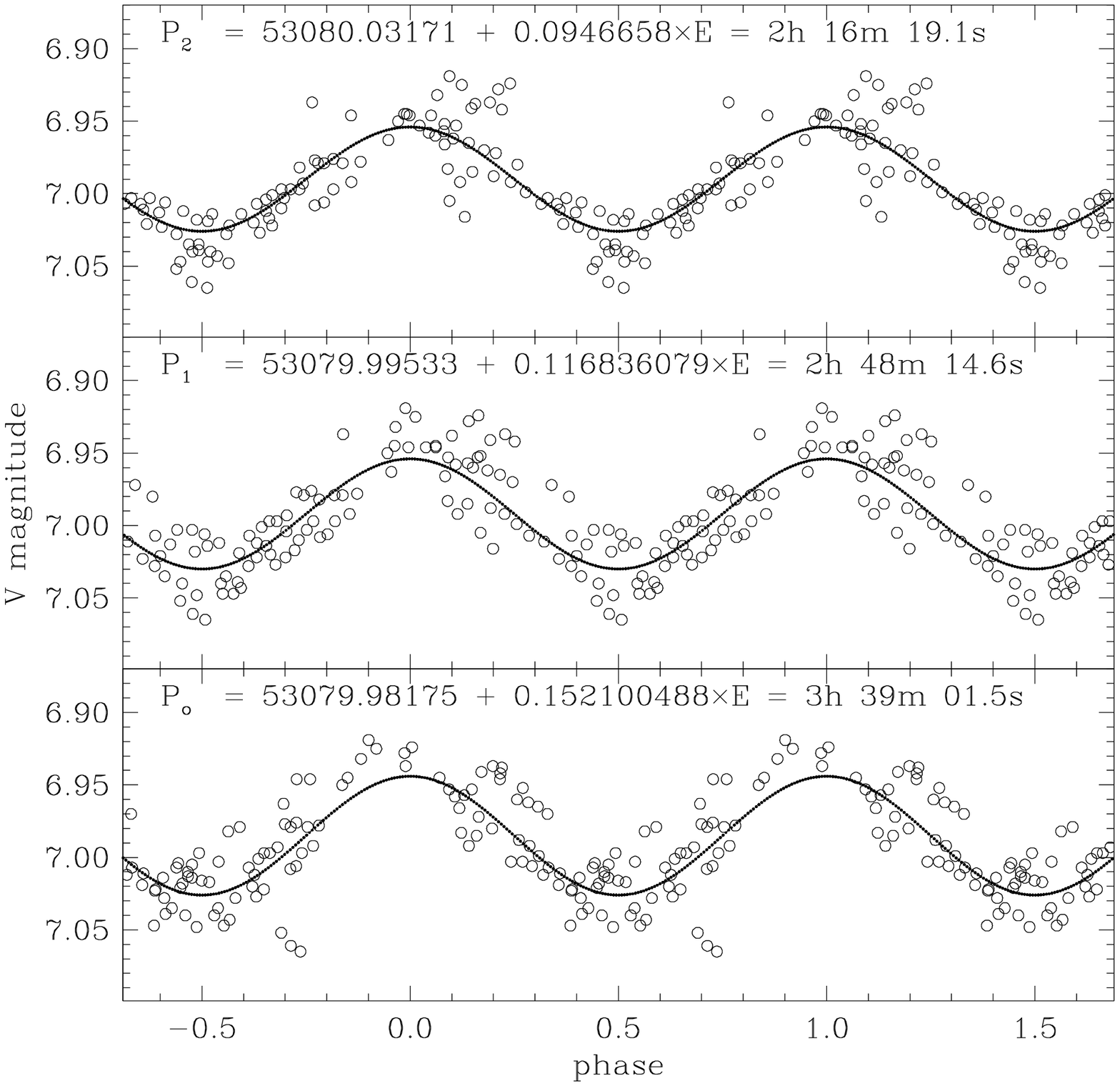}{Our V-band data of V994~Her outside eclipses folded onto the three main
periodicities (fundamental, I and II overtone) derived for the $\delta$-Sct pulsation via Fourier
analysis.}

All together, we have collected 1170 measurements in $V$ band of V994~Her
between June 10, 2002 and July 15, 2004. All observations were corrected
for atmospheric extinction and color equation (via calibration on
Landolt's equatorial fields), and the instrumental differential magnitudes
were transformed into the standard Johnson UBV system. The variable,
comparison and check stars are close on the sky so the atmospheric
corrections were rather small (108 arcmin distance for HD\,168957 and 66
arcmin for HD\,170650). The close similarity of the color between the
variable and comparison star and the fact that all observations have
been obtained for zenith distances $<60^\circ$ argue for a high internal
consistency of our photometry of V994~Her.

A Deeming-Fourier code has been applied to the set of data, resulting in
the following ephemeris for the orbital period of the eclipsing binary
\begin{equation}
{\rm Min.\ I = HJD\ }2452436.4213(\pm 0.0003)\ \  +\ \  2\fday083218(\pm 0.000002) \times E
\end{equation}
Our $V$ photometric data folded to this ephemeris are presented in Figure~1.
Outside eclipses it is $V$=6.91 and at primary minimum V994 Her gets fainter
by $\Delta V$$\sim$0.27~mag. The secondary eclipse falls around orbital
phase 0.52, indicating an eccentric orbit.

\IBVSfig{13.5cm}{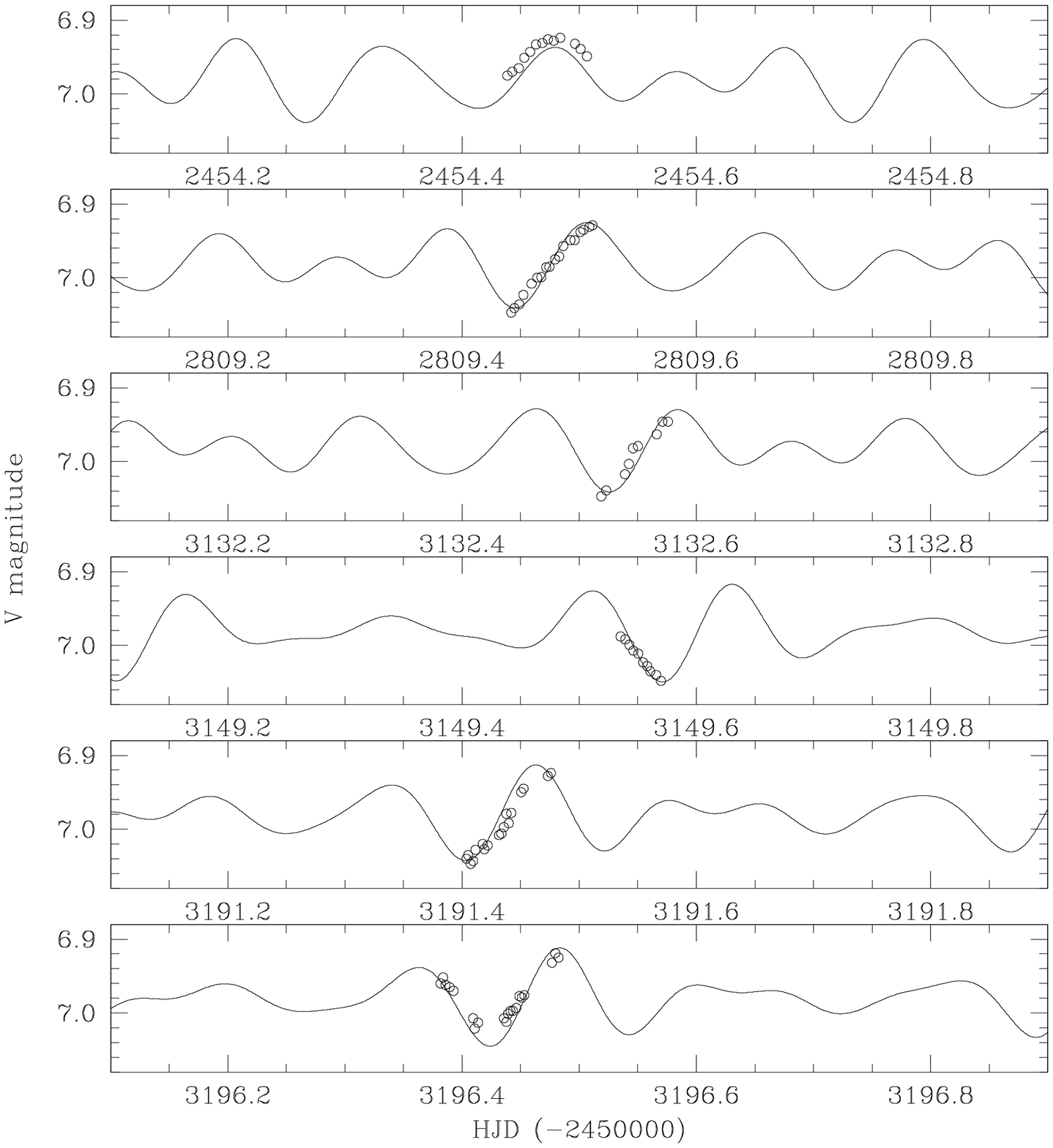}{Sample segments of our V-band data of V994 Her away from eclipses
compared with the multi-mode $\delta$-Sct pulsation as given by Eq. (4).}

However, the most attracting feature of the lightcurve in Figure~1 is the
presence of a strong, short period variability in addition to the eclipse
modulation. The amplitude of such variability is of the order of $\Delta
V$$\sim$0.13 mag, with a period of a few hours. This short period
variability originates in the eclipsing binary, not in the nearby
astrometric companion. In fact, Hipparcos lightcurve for the eclipsing
binary alone show a similar `scatter' of the data outside eclipses.

We have used a Deeming-Fourier code to search for periodicities in the
photometric data of V994~Her outside the eclipses. Three firm periodicities
($0\fday152100$, $0\fday116836$, $0\fday094666$) emerge from the
analysis. The data outside eclipses are phase plotted against each of them in
Figure~2. Their periods and amplitudes are reminiscent of $\delta$~Sct type
of variability. It is worth noticing that their ratios are:
\begin{equation}
\frac{\rm P_1}{\rm P_o}=0.786 \phantom{xxxxxxxxxx} \frac{\rm P_2}{\rm P_1}=0.810
\end{equation}
which are very typical of $\delta$~Sct variables (e.g. Fitch 1976) and
closely similar to those expected from theoretical modeling (e.g. Cox et al.
1979). We are therefore inclined to conclude that in V994~Her there is a
$\delta$~Sct-like variable showing multi-mode pulsations, caused by the
contemporaneous presence of fundamental ($P_o=0\fday152100$), first overtone
($P_1=0\fday116836$) and second overtone ($P_2=0\fday094666$) modes.

Our observations have been typically collected a few per night. However, in
some occasions, longer sequences have been acquired. Those away from eclipse
phases are plotted versus time in Figure~3, where the continuous line is the
multi-mode pulsation described by the sine-wave equation:
\begin{eqnarray}
V&=\ 6.98& +\ 0.031\sin\ \left[ 2\pi \left( \frac{t-53079.9818}{0.152100488}-\frac{1}{4} \right) \right]\\ \nonumber
 &       & +\ 0.025\sin\ \left[ 2\pi \left( \frac{t-53079.9953}{0.116836079}-\frac{1}{4} \right) \right]\\ \nonumber 
 &       & +\ 0.015\sin\ \left[ 2\pi \left( \frac{t-53080.0417}{0.09466580}-\frac{1}{4} \right) \right]
\end{eqnarray}
that provides a good fit to the observed data, indicating the continous
presence - with stable amplitude and phase - of these three modes of
pulsation over the two years spanned by the observations in Figure~3.

\references 

Ak H., Filiz N. 2003, IBVS 5462

Abt H.A. 1985, ApJS 59, 95

Bessell, M.S. 2000, PASP 112, 961

Borkovits T. et al. 2002, IBVS 5313

Borkovits T. et al. 2004, IBVS 5579

Cox A.N. et al. 1979, ApJ 228, 870

Fabricius, C., Makarov, V.V. 2000, A\&A 356, 141

Fitch W.S. 1976, IAU Coll 29, 185

Ling J.F., Prieto C. 2000, A\&AS 143, 335

Munari U. et al.  2004, A\&A 418, L31

Siviero A. et al.   2004, A\&A 417, 1083

\endreferences 

\end{document}